%% file: cramer2.tex
\documentclass{article}
\usepackage{bezier,emlines2,gdgspace}
\advance\textheight by -\headsep\advance\textheight by -\headheight
\begin{document}
\doublespacing
\centerline{\Large Cramer's Transactional Interpretation
and Causal Loop Problems}\vskip 1cm
\begin{center}
{\normalsize \em Ruth E. Kastner \\ Department of Philosophy \\
University of Maryland \\
College Park, MD 20742 USA. \\
July 9, 2004}
\end{center}

\vskip 1cm

\vskip 1cm\large

ABSTRACT. Tim Maudlin's argument for the inconsistency of
Cramer's Transactional Interpretation (TI) of quantum theory has been
considered in some detail by Joseph Berkovitz, who has
provided a possible solution to this challenge at the
cost of a significant empirical lacuna on the part of TI.
The present paper proposes an alternative solution in 
which Maudlin's charge of inconsistency is evaded but
at no cost of empirical content on the part of TI. However,
Maudlin's argument is taken as ruling out Cramer's heuristic
``pseudotime'' explanation of the realization of one transaction out of
many.

\vskip .7cm

1. Introduction

John Cramer's Transactional Interpretation (TI) 
(Cramer, 1986) is based on
the Wheeler-Feynman emitter-absorber theory of radiation
(Wheeler and Feynman 1945),
which proposed that electromagnetic interactions involve
time-reversed ``advanced waves'' as well as the usual
retarded solutions to the wave equation. Cramer extended
the notion of advanced wave solutions to the quantum
domain, proposing that in any quantum mechanical
interaction, both types of waves are present. That is,
he proposed that the complex conjugate of the Schrodinger
equation and its solution, $\psi^*$, play an equal
role with the usual Schrodinger equation, and that the latter
represents an advanced wave while the usual solution $\psi$,
represents the retarded wave. He used these premises
to derive an elegant ontological account 
of the relationship of quantum mechanical amplitudes
to observable probabilities (i.e., the Born rule),
as well as a proposed resolution of the problem of 
wave-function ``collapse'' and other puzzles. 

 In TI, a quantum system is produced by a source S
which plays the role of an ``emitter'' in the
Wheeler-Feynman theory. However, unlike in standard quantum theory,
the source emits both a quantum mechanical wave $\psi$
and a time-reversed counterpart, $\psi^*$, exactly out of
phase with $\psi$.

Cramer refers to the future-directed $\psi$ -wave
as an ``offer wave'' (OW). This wave continues on until it
interacts with an absorber, which absorbs the wave
and in response emits a ``confirmation wave'' (CW) also having
two components, both advanced and retarded. Offer waves and 
confirmation waves extending forward in time beyond
the absorber and backward in time beyond the emitter are
exactly out of phase. If the absorber's returned confirmation wave
is equal in amplitude to the offer wave, the ``pre-emission''
waves and the ``post-absorption'' wave mutually cancel; the only nonzero
field that remains
is on the worldline connecting the source and
the absorber, where a retarded OW and an advanced CW
``overlap.'' The final amplitude of this standing wave is
$\psi^* \psi$, which reflects the Born
probabibility in an elegant manner. 

After the advanced CW reaches
S, there is a possibility for a transaction to occur, which
according to Cramer involves an ``echoing'' process in 
which the emitter and the absorber interaction cyclically
repeats. At some point S ``accepts'' or reinforces the CW 
resulting in a realized transaction and an actual, observable event (such as the detection
of a particle), and which reflects the Born probability as an intrinsic
dynamical feature. Thus, under TI it is only as a result of completed 
transactions that observable, empirically verifiable events
can be said to occur. Quantum particles such as electrons are thus to be identified
with the transactions that give them detectable reality;
in some sense, an electron does not fully exist until a transaction
has occurred.

2. Maudlin's thought experiment

Tim Maudlin (2002)
has presented the following challenge to TI (Refer to Figure 1):
\vskip 1.5cm
\input mfig.tex

At an initial time $t_0$, a source emits particles either to
the left or right, in the quantum state

$$|\psi\rangle = {1\over \sqrt 2}({|L\rangle + |R\rangle})\eqno(1)$$

\noindent with obvious notation. To the right is a detector ``A''
1 unit away and behind it another detector ``B'' 2 units away,
also initially on the right but with the capability of being
swung quickly over to the left.
A may detect a particle with a probability of $1\over 2$.
If A does not detect the particle after the requisite time interval
has passed, then, at time $t_1$, B is swung quickly around to the left to
 detect the particle. To avoid ambiguities and practical
problems arising from experimenter intervention, we can
assume that B, initially located on the right behind A,
is equipped with a timed circuit that will automatically cause it to
be swung to the left after the appropriate
time interval, unless it receives an `abort' signal
from the detection of the particle at A. So the detection
process is fully automated.

According to TI, offer waves of amplitude ${1\over \sqrt 2}$
are emitted to the right and to the left. Detector A returns a
confirmation wave of amplitude ${1\over \sqrt 2}$ but detector
B cannot return a confirmation wave since it is being blocked by A.
However, if no realized transaction occurs between the source and A,
then B is swung over to the left and then it returns a confirmation
wave of amplitude ${1\over \sqrt 2}$, while detecting the particle with
certainty.

Maudlin argues that, since whenever a confirmation wave
is returned from B, the particle is certain to be detected
there, it is inconsistent for the amplitude of that
confirmation wave to be only ${1\over \sqrt 2}$, since that
implies a probability of only $1\over 2$ for detection
of the particle at B. He concludes,
based on this argument, that ``Cramer's theory collapses.''

Besides this apparent inconsistency between predicted frequency
of detection at B being unity whenever B is in place
and the intrinsic probability of $1\over 2$
based on the L-confirmation wave,
there is another difficulty raised by this thought experiment.
This second difficulty concerns the role of what Cramer terms
a ``pseudotime'' cyclic sequence, in which the particular realized transaction
is supposed to result from an echoing back and forth of
offer and confirmation waves between the emitter and the
absorbers, until an outcome occurs in which the appropriate
conservation laws 
(such as energy, momentum, etc.) are satisfied. As Maudlin
points out, such an ``echoing'' process is reminiscent of
the usual way of thinking about interior field values being determined
as a result of fixed boundary conditions. But in this example,
the boundary conditions are not fixed, but are instead
causally dependent on the very outcomes that---according
to Cramer's presentation of his theory---are supposed to
be determined, at least in part, by those boundary conditions.

So the Maudlin challenge to TI can be characterized by two
distinct aspects: (1) the apparent inconsistency between
the predicted frequency and the intrinsic probability of
detection at L; and (2) the apparent inadequacy of the TI
description of the realized transaction as resulting from
fixed and independent boundary conditions,
in the face of a realizable experimental situation in which
the boundary conditions are not fixed and independent.

Concerning (2), the pseudotime narrative talks about
a sequence of events in which (i) an offer wave is emitted,
(ii) various absorbers return confirmation waves to the
emitter, and then (iii) the emitter responds to the confirmation
waves, with the process repeating (iv) until a transaction
is realized. Thus, at least as seen in pseudotime,
the transaction and corresponding outcome does not
occur until step (iv). But in this experiment, the
nature of the particular confirmation waves arising in
step (ii) is dependent on the outcome of the
experiment---that is, what is supposed to be as yet undecided
until step (iv). So the process
of ``echoing'' leading to the realized transaction
cannot include the left-hand component
of the offer wave, which will only result in a confirmation wave
based on a definite outcome of ``particle not on the right'' having already occurred,
presumably at step (ii). Therefore, if it turns out that
the particle is detected on the left, it can't be because
of any echoing process between the emitter and absorber B.

This shows that there is something wrong with the pseudotime
narrative which asserts that the realized transaction does not
occur until the ``echoing'' process of step (iv). So,
whether or not Maudlin's argument against the probabilities
ultimately holds up, it seems clear that the picture of 
the realized transaction coming about through a cyclic
echoing process, based on fixed boundary conditions, is suspect.

However, arguably the heuristic ``pseudotime'' account is not
crucial to Cramer's interpretation. In fact the core of TI
is the assertion that advanced waves from absorbers play an important and heretofore 
neglected role as the ontological basis of the probability of 
an outcome. In this paper I take the point of view that the
``pseudotime'' account, however discredited by the Maudlin
thought experiment, is not fundamentally constitutive of TI and can
be discarded without significant harm to the basic interpretation.
It is assumed here that the main thrust of TI is that
the relationship between an emitter and a set of absorbers
``carves up'' the probability space corresponding to
the possible outcomes defined by that configuration. Specifically,
it is assumed that the weights of possible transactions
define a partitioning of the probability space. More details
of this picture will be provided in section 4.

3. Berkovitz' account of frequencies in causal loops

J. Berkovitz (2002) offers an analysis of Maudlin's problem
in terms of different interpretations of the probabilities involved,
and concludes that Cramer's TI can evade the charge of
inconsistency given the appropriate interpretation of the
probabilities prescribed by TI. The argument essentially consists
of the claim that Maudlin's experiment constitutes a causal loop,
and that intrinsic probabilities (such as those suggested
by the initial conditions of the experiment in terms
of the offer waves and confirmation waves in place
at $t_0$) cannot be expected to equal (or be close to)
the long-run frequencies in causal loops.

Berkovitz' basic illustration of an indeterministic causal
loop is that of a coin toss: A balanced coin is tossed (A),
which indeterministically causes the result ``heads'' (B),
which deterministically causes my perception of ``heads'' (C),
which deterministically and in the reverse time direction
causes the coin flip (A). While a coin toss outside of
a loop would have a long-run frequency of 1/2 for ``heads,''
in this loop that frequency is 1 due to the constraints of
the loop: whenever the coin toss occurs, we know that
outcome ``heads'' occurs.

Berkovitz argues that such a loop can be seen as consistent
by noting that the reference class of causal states (A) giving
rise to (B) is biased due to the reverse-time deterministic connection between
(C) and (A). That is, whether or not the cause (A) occurs is
not independent of its effect (B).
Therefore, the long-run frequency of
(B) should not be expected to correspond to its unbiased
probability in the reference class of (A) and the discrepancy between the
two is not an indication of inconsistency. 
Berkovitz also notes that the conditional
probability of B given A in the causal circumstances of
the loop is unity, i.e., $P(B|A) = 1$, which is equal
to the long run frequency. Either of these facts can be
seen as showing the consistency of the loop, i.e. that
such a loop is physically possible.

To be more precise about Berkovitz'argument we need to
briefly review the formulation he presents, which takes off from
Butterfield's proposal to distinguish between ``many-spaces''
and ``big-space'' probabilities (Butterfield 1989, 1992).
Consider an experiment with possible states $\lambda_i$ and 
measurement settings $S_j$ and
possible outcomes X or Y. The ``many-spaces'' approach
assigns a different probability space to each of the possible
experimental conditions defined by the states and
settings, with the probabilities for each outcome defined only
within that space. Butterfield translates this approach in
logical terms as

`conditional with a probabilistic consequent': $(S_j \& \lambda_i) \rightarrow(pr(X)=x)$

where $x$ is a number between 0 and 1.

The notation for the many-spaces probability of outcome X in
the reference class of state $\lambda_k$ and setting $S_l$ is
$P_{\lambda_k, S_l} (X)$. In the case of the coin-toss loop
(Berkovitz' ``Loop 1''), the many-spaces probability of ``heads'' (B) on 
a balanced coin toss (A) is written as $P_{A} (B) = 1/2$.

In contrast, the ``big-space'' approach uses a single probability
space to assign probabilities to outcomes with reference
to particular states and settings, as conditional probabilities
defined in the usual way (as the conjunction
of all the events divided by the probability of the settings
and states). Thus the ``big-space'' probability of X for the
same state and setting as above is written
$P(X|\lambda_k, S_l)$.

The crucial conceptual difference between these two approaches
is the following. In the big-space approach, absolute probabilities
must be defined for all possible states and settings. 
In contrast, the many-spaces
approach treats states and measurement settings as ``exogenous'' variables
whose probabilities are undefined. If a particular experiment
involving setting $S_j$ is being performed, then
probabilities of outcomes for a different 
(counterfactual) setting $S_k$ are treated
as subjunctive probabilities. Thus in this sense, the ``many-spaces''
approach treats the different spaces as separate possible worlds
whose intrinsic probabilities are undefined and irrelevant
to the question being asked. Butterfield (1992, p. 47) motivates this picture
in the following way:

``The big space is committed to probabilities for acts of measurement,
which the many space construal avoids...for an act of measurement
surely need not have a probability. Why should every proposition
or event have a probability? And since $a$ is a feature of a complex
apparatus, and is fixed or at least influenced by the choice of
the experimenter, it seems a good candidate for not having
a probability.''

Berkovitz provides persuasive arguments for the ``many-spaces'' approach
in the context of Bell-type experiments testing a relativistic
parameter-dependent (PD) hidden variable theory. While Arntzenius (1994) has
presented an inconsistency proof for such theories, Berkovitz
argues that the proof is dependent on a big-space approach
which is inappropriate in that it often reflects 
the specific experimental setup rather than 
the structure of the theory under consideration. He
argues that a many-spaces approach blocks the proof
because it involves a causal loop in which
the many-space probabilities should not be expected to equal
either the long-run frequencies or conditional probabilities
associated with the loop. 

Berkovitz bolsters his argument against the applicability
of big-space probabilities in the context of Bell-type
experiments by showing  that using the ``big-space'' approach leads to an absurd
conclusion that Bohm's theory is inconsistent. 
However, while offering relativistic PD theories a way out of Arntzenius'
impossibility proof,
Berkovitz notes that a further difficulty faced by such theories
is a lacuna in their prediction of the ``unconditional'' frequencies of outcomes, 
in the face of the causal loops they create. In
a many-spaces approach, the loops will result
in an apparently unspecifiable deviation of the long-run frequencies from the many-spaces
probabilities of outcomes. Also, the big-space approach not only
runs afoul of the proof but fails to pin down unique long-run frequencies
because the constraints of the loop are too weak.

4. Berkovitz' solution, its price, and an alternative

Berkovitz' suggested evasion of the Maudlin confirmation-wave
inconsistency is for TI to define the probability
for detection at B (on the left) 
as the many-spaces probability $P_{\psi}(L)$ and point out that the conditional
frequency $f(L|\psi)$ should not be expected to be equal
to this probability in a causal loop. This is a legitimate
approach, but as in the relativistic PD theory case, 
many-spaces probabilities applied to causal loops
will fail to provide unambiguous predictions for
long-run unconditional frequencies of specific outcomes
such as detection on the left (L) or on the right (R).
  
It will be argued below that in fact Cramer's theory can provide
unconditional frequencies for L and R outcomes if one
abandons the account of a pseudotime ``echoing'' among non-fixed
absorbers, and interprets Cramer's theory
 as providing for an unambiguous partitioning
of a ``big'' probability space based on the intrinsic
weight of the possible transactions. It will
also be necessary to adopt a fully time-symmetric account of causal
dependence, which, if somewhat radical, can be seen as
consistent with the explicit time-symmetry of TI.

Firstly, it should be noted that, under Cramer's theory, it is
not necessary to have a complete set of absorbers (that is,
an absorber for every possible outcome of a complete
set of observables) in order to define a complete set of
definite outcomes. For example, consider the trivial
version of Maudlin's experiment in which there is no detector
B at all. Only one possible transaction can be formed, as
there will only be a confirmation wave returned from A.
But the probability that this transaction will be realized
is still only 1/2: the particle may or may not be detected
at A. Thus the two possible outcomes in this experiment are
``particle detected at A,'' denoted $R_d$ and ``particle not detected at A,''
denoted $\neg(R_d)$, each with a probability of 1/2. 
The latter ``null'' outcome corresponds to there being
no transaction formed, always a possibility in Cramer's theory
if there are not absorbers present for every possible
eigenvalue of an observable. 

Recall also that under TI an emitter emits both the usual
retarded wave and also an advanced wave which propagates
in the direction of decreasing time index. According to TI,
this advanced wave is exactly out of phase with 
any confirmation waves returned
from absorbers. Thus when there is a complete set of
absorbers (i.e., when the sum of the confirmation ``echoes''
from all the absorbers is equal in amplitude to that of
the offer wave), the advanced wave from the emitter
is exactly canceled. However, when the set of
absorbers is not complete, the emitter's advanced wave
is not completely canceled.

In the single-absorber case discussed above, there is only
a confirmation wave ${1\over \sqrt 2} \langle R|$ returned from A,
which when added to the emitter advanced wave, 
$-{1\over \sqrt 2} [\langle L| + \langle R|]$, leaves
a remaining emitter advanced wave of 
$-{1\over \sqrt 2} \langle L|$. 

Returning to the two possible loops in Maudlin's experiment,
we see that the loop based on detection at A will be accompanied
by a remnant of emitter advanced wave as calculated above,
while the loop based on detection at B will be accompanied
by zero emitter advanced wave, owing to the fact that
there is a complete set of absorbers and therefore complete
cancellation of the emitter's advanced wave as discussed above.

In the usual application of TI, the set of absorbers 
(whether complete or incomplete) is fixed throughout the experiment,
as is the emitter state (including 
confirmation waves arriving back at the emitter from any 
absorbers).
One can therefore think of the emitter state at the
time of emission as the single ``branch'' event having several possible
futures, say indexed by $i$. It is then easy to think of the past (relative to the
branch event) as fixed and
the future (again, relative to the branch event) as indeterminate, 
in the usual time-asymmetric
way, and assign to these possible futures the ``unconditional''
probabilities or frequencies corresponding to the weight
of the corresponding transaction ($\psi^*_i \psi_i$). 

However, in the Maudlin example the past, up to and even
including the emission event, is not fixed. We therefore cannot think of
the emission event as a ``branch'' point and cannot define
unconditional probabilities for the two outcomes in the
usual way. Hence Berkovitz' claim that it appears to be
impossible to calculate unconditional frequencies for the
outcomes L and R. Nevertheless, it is proposed here that
since TI is a fully time-symmetric interpretation and moreover
since that time symmetry is reflected in causal effects
``radiating,'' as it were, out in both temporal directions
from the loop's two
possible outcomes (as discussed above in terms of the 
varying emitter advanced waves), that this problem can be
solved by attacking the problem from a time-symmetric
standpoint. This means that we should expect neither the
past nor the future (relative to the branching event) to be fixed.
But we also have to identify a different branching event
than the emission state at $t_0$, since as Berkovitz
points out, this state itself depends on the outcome
and is therefore not independent.

The correct time-symmetric branch point will
be found by identifying the event which is shared by both loops
(just as the emission event is shared by both branches
in the usual time-asymmetric situation). It is what both
loops have in common: the overlap of the offer and
confirmation waves corresponding to R, i.e., the
field $\psi_R + \psi_R^*$ (where $\psi_R$ denotes the
component of the offer wave absorbed by R), between $t_0$ and $t_1$. In both loops this
field exists; however in one of the loops (R) it becomes
a realized transaction and in the other (L) it becomes
a failed transaction. There is no way to predict whether
this field will end up as a detected particle or not
(and absent the ``pseudotime'' account we don't as yet
even have a heuristic way to understand this process),
but TI provides for the probability of each 
outcome---1/2---and this is precisely the unconditional frequency
of each outcome R and not-R, if we view this event as the
branch point.

With this probability assignment, we can define a big-space probability
for the Maudlin experiment as follows (see Figure 2.)
Maudlin's experiment provides for two possible
measurement processes, each of which is deterministically dependent
on the outcomes, each of which has an unconditional
probability of 1/2. As discussed above, the minimal absorber arrangement
allows us to define two basic outcomes, R-emission/detection 
$R_e = R_d$, and $\neg(R_d)$. When
B is on the left then $\neg(R_d) = L_d$.
The big probability space is therefore divided in half
according to these possible outcomes, with the probability
of each being 1/2. Now Maudlin's experiment
dictates the probabilities of measurement setting
according to these possible outcomes in the following
way: the region associated with
$\neg(R_d)$ corresponds to the probability of measurement of both R and L;
yet in this region the outcome is of course known to be $L_d$.
The other region associated with $R_d$ corresponds to the probability
of measurement of R only.
Therefore, the ``augmented'' initial states of
the emitter are labeled (as in Berkovitz 2002) in each respective region of the
big probability space by $\psi_C$ and $\psi^{\prime}_C$,
where the former includes confirmation waves from
A and B and the latter includes only a confirmation wave
from A.\vskip 1.7cm

\input prob.tex
\newpage
We can now obtain conditional big-space probabilities
so as to enable Cramer to escape from the trap, in the
following way:

$$P(L_d|\psi_C) = {P(L_d \& \psi_C)\over P(\psi_C)}
= {(1/2)\over (1/2)} = 1\eqno(2)$$

Referring again to Figure 2, the two equal portions of
the big probability space can be intuitively thought of
as two distinct possible worlds created by the minimal
emitter/absorber configuration. The incipient transaction
corresponding to the field $\psi_R + \psi_R^*$ can be thought of as
an unstable ``bifurcation line'' between the two worlds.
When that transaction succeeds, the system enters
the right-hand region; when it fails, the system enters
the left-hand region. Since in the latter case B swings over to
the left and emits a confirmation wave $\Psi_L^*$, what would have otherwise
been a null outcome becomes a realized transaction resulting
in $L_d$. (Note that this account is only possible if we
abandon the idea that there is cyclic ``echoing'' between
B and the emitter if such echoing is taken as reflective of
an uncertainty in outcome.
For the measurement of L-emission takes place only
when the outcome is already certain; the system has
already entered the left-hand region of the probability
space.)

The fact that the amplitude of the confirmation wave from
B is only $1\over \sqrt 2$ shows that confirmation waves 
are properly interpreted
as reflecting the entire big-space probability structure: despite the
fact that when B is moved to left the outcome is already
manifest, the confirmation
wave still ``knows'' that the particle will only be detected
at B in 1/2 of the trials---that only half the offer wave is directed toward the left.
It thus retains the full information corresponding
to the set of {\it both} loops, and therefore must not be a property of 
only $\psi_c$ but the entire experimental arrangement which
contains the possibility of $\psi^{\prime}_c$ as well. This
point is perfectly in keeping with the well-known phenomenon
of ``quantum wholeness'' and should therefore not be
entirely unexpected.

5. Is there a bilking problem due to the emitter advanced wave?

There might appear to be a possible snag with this proposed solution.
As discussed above, the emitter advanced wave differs for each loop. 
This means that events {\it prior} to the spacetime point of the emission  
differ for each outcome/loop. That is, suppose it happens that outcome/loop $R_d$
occurs at $t_0$. Then there exists a nonzero emitter advanced wave for
all times $t < t_0$ and the future outcome from the standpoint of
any of these earlier times is apparently already decided. Might this give
rise to a bilking problem, i.e., could a contradiction be arranged
wherein a different outcome is brought about?

The answer is ``no,'' because there is no way to detect the
existence of the nonzero emitter advanced wave. That is, in 
order to create a bilking problem one would need to
discover the nature of the emitter advanced wave, and then
arrange it so that that advanced wave could never be
produced, presumably by cancelling the proposed experiment,
or modifying it appropriately. But there doesn't appear
to be any way to detect this remnant of offer wave
coming from the future; in order to somehow engage
it in a transaction (which is the way things are detected
in TI), a retarded offer wave would
have to be perfectly in phase with it.

Thus, relative to a time prior to $t_0$, even though the world they
inhabit will in some sense ``already'' bear the imprint of the future outcome
of the Maudlin experiment--whether R or L--the human experimenters
have no access to that
information, and in fact it isn't even ``actual'' for them.
(Recall that under TI ``actualized'' events,  or empirical facts,
result only from realized transactions). So from their perspective,
the result of the experiment is still uncertain, even though
the causal effects of the experiment radiate out in both
temporal directions.

Huw Price explores the issue of advanced causal effects
(what he terms ``advanced action'') and their relationship to the bilking problem
in detail in his (1996), Chapter 7. He argues (similarly to 
Michael Dummett (1954, 1964) decades earlier) that it is consistent
to assert that an effect can precede its cause provided that
the claimed earlier effect is epistemically inaccessible
to anyone who might try to set up a bilking problem as
described above. Price concludes (in arguing for the
time symmetry of counterfactual dependence) that the past should not
be considered to be ``cast in stone'' and that questions
about what we can affect are properly answered in terms
of epistemological accessibility, not temporal relationships:

``Even if experience teaches us that whatever we know about via
memory and the senses lies in the past, this does not imply
that anything that lies in the past is something that might in
principle be known about...In fact, it seems that the relationship
between temporal location and epistemological accessibility
is not only contingent (in both directions), but rather
underdetermined by our actual experience.'' (Price 1996, p. 175)

In these terms, it is coherent to claim that the incipient transaction
between the emitter and A (i.e., the field $\psi_R^* + \psi_R$)
is the cause of various effects
that lie not only in the future of the emission event but
also in its past. As an independent cause of these effects,
the incipient transaction can legitimately be considered
the independent ``branch point'' and its probabilities
of occurrence and non-occurrence can be considered the
independent probabilities needed to provide the
unconditional frequencies of the outcomes L and R.

6. Conclusion.

 Berkovitz' proposed escape route for
Cramer's Transactional Interpretation
 from Maudlin's causal loop inconsistency claim requires assuming
a many-spaces approach to probabilities together with an argument that
the deviation of long-run frequencies from those probabilities
is unproblematic. However, that solution comes with the
price that TI can make no empirical prediction for the frequencies
of the two possible outcomes. 

An alternative solution has been proposed, based on a big-space
approach to probability which applies the explicit time-symmetry
of TI to the causal dependence of events, in order to define
unconditional frequencies of outcomes.
The transactional event
which is common to both possible loops in the Maudlin example
is treated as the independent ``branch point'' relative
to which independent probabilities, determined by the
weights of the respective transactions, can be assigned.

It is argued that Maudlin's thought experiment shows that
the pseudotime ``echoing'' account of the realized transaction
is flawed; the big-space approach
presented here depends on abandoning that pseudotime account
which appears to require that the realization of a transaction
depends on all possible absorbers, whether fixed or not. 
Nevertheless, it is argued that the pseudotime narrative
is merely heuristic and is not a crucial part of Cramer's
theory.

 The big-space probability reflects the
fact that the specific measurement settings and states
are governed by a clearly defined
probability structure, as opposed to a many-spaces approach
in which these quantities are arbitrary or undefined.
 It also provides a natural explanation
of the puzzling feature that the amplitude of the confirmation wave from B
is only ${1\over\sqrt 2}$, in terms of ``quantum wholeness'':
the confirmation wave retains information about the entire
experimental arrangement.
\vskip 1.5cm
Acknowledgements

The author is indebted to Joseph Berkovitz, Tim
Maudlin, and David Miller for valuable discussions and correspondence.
\newpage

References.

\noindent Arntzenius, Frank (1994), ``Space-Like Connections,'' {\it British Journal
for the Philosophy of Science} 45: 201-217.\newline
Berkovitz, Joseph (2002), ``On Causal Loops in the Quantum Realm,''
in T. Placek and J. Butterfield (eds.), {\it Non-locality
    and Modality}, Proceedings of the NATO Advanced Research Workshop on Modality,
Probability
    and Bell's Theorems, Kluwer, pp. 233-255.\newline
Butterfield, Jeremy (1989), ``A Space-Time Approach to the Bell
Inequality,'' in Cushing and McMullin, eds. (1989). pp. 114-144.\newline
Butterfield, Jeremy (1992), ``Bell's Theorem: What It Takes,''
{\it British Journal for the Philosophy of Science} 42: 41-83.\newline
Cramer, John G. (1980), ``Generalized Absorber Theory
and the Einstein-Podolsky-Rosen Paradox,'' {\it Phys. Rev. D} 22: 362-376.
\newline
Cramer, John G. (1986), ``The Transactional Interpretation
of Quantum Mechanics, {\it Rev. Mod. Phys.} 58: 647-688. \newline
Cushing, J. and E. McMullin, eds. (1989),
{\it Philosophical Consequences of Quantum Theory:
Reflections on Bell's Theorem}, Notre Dame: University of Notre
Dame Press.\newline
Dummett, Michael (1954), ``Can an Effect Precede Its Cause?,'' 
{\it Proceedings of the Aristotelian Society, Supplementary
Volume} 38: 27-44.\newline
Dummett, Michael (1964), ``Bringing about the Past,''
{\it Philosophical Review} 73: 338-359.\newline
Maudlin, Tim (2002), {\it Quantum Non-Locality and Relativity},
Second Edition, Oxford: Blackwell.\newline
Price, Huw (1996), {\it Time's Arrow and Archimedes' Point}, Oxford: Oxford 
University Press.\newline
Wheeler, J. A. and R. P. Feynman (1945), ``Interaction with the Absorber
as the Mechanism of Radiation,'' 
{\it Reviews of Modern Physics} 17: 157-181.
\end{document}

%% file: mfig.tex
\special{em:linewidth 0.4pt}
\unitlength 1.00mm
\linethickness{0.4pt}
\begin{picture}(154.67,129.00)
\put(55.67,40.00){\framebox(39.67,20.00)[cc]{$\psi$}}
\put(122.33,50.00){\vector(1,0){0.2}}
\emline{95.33}{50.00}{1}{122.33}{50.00}{2}
\put(123.00,40.00){\rule{2.00\unitlength}{20.00\unitlength}}
\put(1.67,39.67){\dashbox{2.67}(2.33,20.33)[cc]{ }}
\put(29.33,50.00){\vector(-1,0){0.2}}
\emline{55.67}{50.00}{3}{29.33}{50.00}{4}
\put(2.67,62.67){\vector(-4,-3){0.2}}
\emline{152.67}{60.33}{5}{149.00}{63.67}{6}
\emline{149.00}{63.67}{7}{145.32}{66.84}{8}
\emline{145.32}{66.84}{9}{141.65}{69.84}{10}
\emline{141.65}{69.84}{11}{137.96}{72.67}{12}
\emline{137.96}{72.67}{13}{134.28}{75.34}{14}
\emline{134.28}{75.34}{15}{130.59}{77.84}{16}
\emline{130.59}{77.84}{17}{126.90}{80.17}{18}
\emline{126.90}{80.17}{19}{123.21}{82.34}{20}
\emline{123.21}{82.34}{21}{119.51}{84.33}{22}
\emline{119.51}{84.33}{23}{115.81}{86.16}{24}
\emline{115.81}{86.16}{25}{112.10}{87.82}{26}
\emline{112.10}{87.82}{27}{108.40}{89.32}{28}
\emline{108.40}{89.32}{29}{104.68}{90.65}{30}
\emline{104.68}{90.65}{31}{100.97}{91.81}{32}
\emline{100.97}{91.81}{33}{97.25}{92.80}{34}
\emline{97.25}{92.80}{35}{93.53}{93.62}{36}
\emline{93.53}{93.62}{37}{89.81}{94.28}{38}
\emline{89.81}{94.28}{39}{86.08}{94.77}{40}
\emline{86.08}{94.77}{41}{82.35}{95.09}{42}
\emline{82.35}{95.09}{43}{78.61}{95.25}{44}
\emline{78.61}{95.25}{45}{74.88}{95.23}{46}
\emline{74.88}{95.23}{47}{71.14}{95.05}{48}
\emline{71.14}{95.05}{49}{67.39}{94.70}{50}
\emline{67.39}{94.70}{51}{63.65}{94.19}{52}
\emline{63.65}{94.19}{53}{59.89}{93.51}{54}
\emline{59.89}{93.51}{55}{56.14}{92.66}{56}
\emline{56.14}{92.66}{57}{52.38}{91.64}{58}
\emline{52.38}{91.64}{59}{48.62}{90.45}{60}
\emline{48.62}{90.45}{61}{44.86}{89.10}{62}
\emline{44.86}{89.10}{63}{41.09}{87.58}{64}
\emline{41.09}{87.58}{65}{37.32}{85.89}{66}
\emline{37.32}{85.89}{67}{33.55}{84.04}{68}
\emline{33.55}{84.04}{69}{29.77}{82.01}{70}
\emline{29.77}{82.01}{71}{25.99}{79.82}{72}
\emline{25.99}{79.82}{73}{22.21}{77.46}{74}
\emline{22.21}{77.46}{75}{18.42}{74.94}{76}
\emline{18.42}{74.94}{77}{14.63}{72.25}{78}
\emline{14.63}{72.25}{79}{10.84}{69.39}{80}
\emline{10.84}{69.39}{81}{7.04}{66.36}{82}
\emline{7.04}{66.36}{83}{2.67}{62.67}{84}
\put(73.33,127.00){\makebox(0,0)[cc]{Figure 1.}}
\put(124.00,32.33){\makebox(0,0)[cc]{A}}
\put(153.00,32.67){\makebox(0,0)[cc]{B}}
\put(152.67,40.00){\rule{2.00\unitlength}{19.67\unitlength}}
\end{picture}

%% file: prob.tex
\special{em:linewidth 0.4pt}
\unitlength 1mm
\linethickness{0.4pt}
\begin{picture}(109.58,126.67)
\emline{77.00}{109.91}{1}{82.06}{109.52}{2}
\emline{82.06}{109.52}{3}{87.00}{108.34}{4}
\emline{87.00}{108.34}{5}{91.69}{106.41}{6}
\emline{91.69}{106.41}{7}{96.03}{103.78}{8}
\emline{96.03}{103.78}{9}{99.91}{100.50}{10}
\emline{99.91}{100.50}{11}{103.23}{96.66}{12}
\emline{103.23}{96.66}{13}{105.91}{92.35}{14}
\emline{105.91}{92.35}{15}{107.89}{87.68}{16}
\emline{107.89}{87.68}{17}{109.13}{82.76}{18}
\emline{109.13}{82.76}{19}{109.58}{77.70}{20}
\emline{109.58}{77.70}{21}{109.24}{72.64}{22}
\emline{109.24}{72.64}{23}{108.12}{67.69}{24}
\emline{108.12}{67.69}{25}{106.25}{62.97}{26}
\emline{106.25}{62.97}{27}{103.66}{58.60}{28}
\emline{103.66}{58.60}{29}{100.43}{54.69}{30}
\emline{100.43}{54.69}{31}{96.63}{51.33}{32}
\emline{96.63}{51.33}{33}{92.35}{48.59}{34}
\emline{92.35}{48.59}{35}{87.70}{46.56}{36}
\emline{87.70}{46.56}{37}{82.79}{45.27}{38}
\emline{82.79}{45.27}{39}{77.74}{44.76}{40}
\emline{77.74}{44.76}{41}{72.67}{45.04}{42}
\emline{72.67}{45.04}{43}{67.71}{46.11}{44}
\emline{67.71}{46.11}{45}{62.97}{47.93}{46}
\emline{62.97}{47.93}{47}{58.57}{50.46}{48}
\emline{58.57}{50.46}{49}{54.62}{53.65}{50}
\emline{54.62}{53.65}{51}{51.22}{57.41}{52}
\emline{51.22}{57.41}{53}{48.44}{61.66}{54}
\emline{48.44}{61.66}{55}{46.35}{66.29}{56}
\emline{46.35}{66.29}{57}{45.00}{71.18}{58}
\emline{45.00}{71.18}{59}{44.44}{76.23}{60}
\emline{44.44}{76.23}{61}{44.66}{81.30}{62}
\emline{44.66}{81.30}{63}{45.67}{86.27}{64}
\emline{45.67}{86.27}{65}{47.44}{91.03}{66}
\emline{47.44}{91.03}{67}{49.92}{95.45}{68}
\emline{49.92}{95.45}{69}{53.07}{99.44}{70}
\emline{53.07}{99.44}{71}{56.79}{102.89}{72}
\emline{56.79}{102.89}{73}{61.00}{105.72}{74}
\emline{61.00}{105.72}{75}{65.61}{107.86}{76}
\emline{65.61}{107.86}{77}{70.49}{109.26}{78}
\emline{70.49}{109.26}{79}{77.00}{109.91}{80}
\emline{76.67}{110.00}{81}{76.67}{44.33}{82}
\put(56.00,81.00){\makebox(0,0)[cc]{$\psi_C, L_d$}}
\put(91.33,80.33){\makebox(0,0)[cc]{$\psi^\prime_C,  R_d$}}
\put(76.67,126.67){\makebox(0,0)[cc]
{Figure 2. Big probability space for the Maudlin experiment.}}
\end{picture}